\documentclass[fleqn,10pt, twocolumn]{wlscirep}

\usepackage{graphicx}
\usepackage{xcolor}
\usepackage{url} 
\usepackage{hyperref} 
\usepackage{dcolumn}
\usepackage{bm}
\usepackage{multirow}
\usepackage[symbol]{footmisc}
\usepackage{amsmath}
\usepackage{subfigure}

\usepackage{url}
\usepackage{soul}
\newcommand{\onlinecite}[1]{\hspace{-1 ex} \nocite{#1}\citenum{#1}} 

\newcommand{\crmnge}{Cr$_{0.82}$Mn$_{0.18}$Ge}
\newcommand{\tc}{$T_\textrm{C}$}

\title{Observation of magnetic skyrmion lattice in Cr$_{0.82}$Mn$_{0.18}$Ge by small-angle neutron scattering}

\author[1,*]{Victor Ukleev}
\author[2]{Tapas Samanta}
\author[3]{Oleg I. Utesov}
\author[4]{Jonathan S. White}
\author[2,1]{Luana Caron}

\affil[1]{Helmholtz-Zentrum Berlin f\"ur Materialien und Energie, D-13109 Berlin, Germany}
\affil[2]{Department of Physics, Bielefeld University, Bielefeld 33501, Germany}
\affil[3]{Center for Theoretical Physics of Complex Systems, Institute for Basic Science, Daejeon 34126, Republic of Korea}
\affil[4]{Laboratory for Neutron Scattering and Imaging (LNS), PSI Center for Neutron and Muon Sciences, Paul Scherrer Institute, CH-5232 Villigen PSI, Switzerland}

\affil[*]{victor.ukleev@helmholtz-berlin.de}

\date{\today}

\begin{abstract}
Incommensurate magnetic phases in chiral cubic crystals are an established source of topological spin textures such as skyrmion and hedgehog lattices, with potential applications in spintronics and information storage. We report a comprehensive small-angle neutron scattering (SANS) study on the $B20$-type chiral magnet \crmnge, exploring its magnetic phase diagram and confirming the stabilization of a skyrmion lattice under low magnetic fields. Our results reveal a helical ground state with a decreasing pitch from 40\,nm to 35\,nm upon cooling, and a skyrmion phase stable in applied magnetic fields of 10-30\,mT, and over an unusually wide temperature range for chiral magnets of 6 K ($\sim T_\textrm{C}/2 < T < T_\textrm{C}$, $T_\textrm{C}=13$ K). The skyrmion lattice forms a standard two-dimensional hexagonal coordination that can be trained into a single domain, distinguishing it from the three-dimensional hedgehog lattice observed in MnGe-based systems. Additionally, we demonstrate the persistence of a metastable SkL at 2\,K, even at zero field. These findings advance our understanding of magnetic textures in Cr-based $B20$ compounds, highlighting \crmnge~as a promising material for further exploration in topological magnetism.

\end{abstract}

\begin{document}
\flushbottom
\maketitle

\thispagestyle{empty}

\section*{Introduction}

Incommensurate magnetic structures, characterized by helical and cycloidal modulations, have recently captured attention owing to the discovery of novel topological spin textures such as skyrmion and chiral soliton lattices \cite{tokura2017emergent,togawa2013interlayer}. The emergence of a chiral proper-screw helical magnetic ground state is prominent in systems lacking mirror symmetry, facilitating Dzyaloshinskiy–Moriya interaction (DMI) \cite{dzyaloshinsky1958thermodynamic,moriya1960anisotropic}. The interplay between Heisenberg exchange, DMI, anisotropy, thermal magnetic fluctuations, and magnetic fields results in a diverse magnetic phase diagram for chiral cubic crystals, encompassing paramagnetic, helimagnetic, field-induced ferromagnetic, conical, and skyrmion lattice (SkL) phases \cite{bak1980theory,bogdanov1994thermodynamically,bauer2016generic}. Recently, magnetic skyrmions have attracted attention as potential candidates for high-density information carriers for racetrack memories and unconventional computing \cite{fert2017magnetic,song2020skyrmion,psaroudaki2023skyrmion}.

\begin{figure*}
\includegraphics[width=1\linewidth]{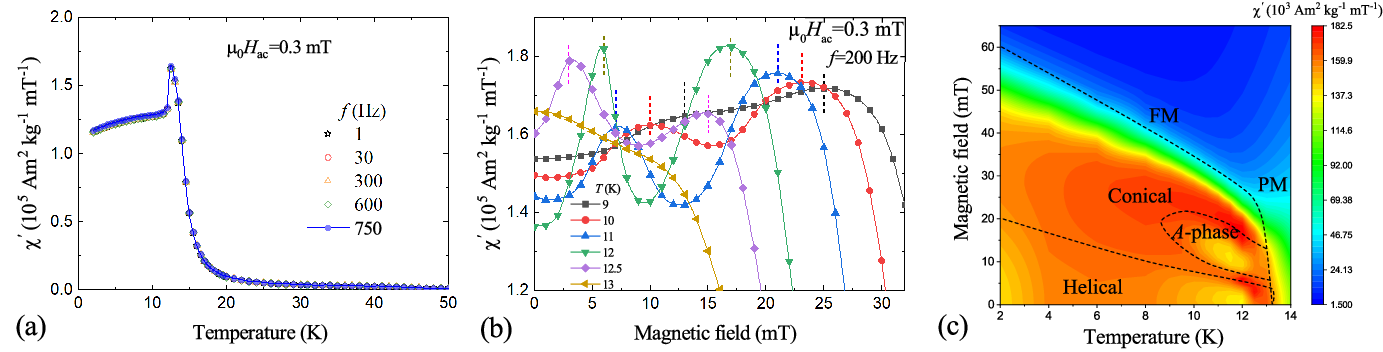}
        \caption{(a) Frequency-dependent ac-susceptibility $\chi'$ measurement. (b) Temperature and field dependence of the $\chi'$ signal around the $A-$phase. Dashed lines indicate the transition fields between the $A-$phase and neighboring conical phases. (c) Magnetic phase diagram constructed from the ac-susceptibility experiment.}
        \label{fig0}
\end{figure*}

Several $B20$-type cubic chiral magnets (space group $P2_13$) \cite{bauer2016generic,tokura2020magnetic} have been identified as hosts for Bloch-type skyrmions, including MnSi \cite{muhlbauer2009skyrmion}, FeCoSi \cite{munzer2010skyrmion}, and FeGe \cite{moskvin2013complex}, with helical and SkL periods ranging from a few tens to a few hundreds of nanometers. Notably, MnGe and solid solutions Mn(Ge$_x$Si$_{1-x}$) in the range $0.2 \le x \le 1$ exhibit complex three-dimensional topological Hedgehog Lattice (HL) textures with a periodicity of a few nanometers \cite{kanazawa2012possible,tanigaki2015real,fujishiro2019topological,kanazawa2020direct,repicky2021atomic,pomjakushin2023topological}. While the SkL is typically confined to the so-called $A$-phase just below the critical temperature \tc~in prototype $B20$ materials \cite{muhlbauer2009skyrmion}, the HL spans the entire magnetic field ($H$) vs. temperature ($T$) phase diagram of MnGe, including low temperatures and high fields up to 25 T \cite{kanazawa2011large,fujishiro2019topological,kanazawa2020direct}. The mechanism of HL formation in Mn-based germanides remains debated due to the vanishing DMI in these systems \cite{gayles2015dzyaloshinskii,bornemann2019complex,grytsiuk2020topological}. Conversely, the metallic paramagnet CrGe, a $B20$ compound, demonstrates ferromagnetic correlations but does not exhibit ordering down to low temperatures \cite{sato1983magnetic,sato1986nearly,klotz2019electronic}.

Interestingly, solid solutions Cr$_{x}$Mn$_{1-x}$Ge present a rich phase diagram, including paramagnetic, ferromagnetic, spin-glass, and helimagnetic regions \cite{sato1988magnetic}. Sato and Morita reported that Cr$_{0.81}$Mn$_{0.19}$Ge features all these phases in the temperature vs. magnetic field phase diagram \cite{sato1999magnetic}. Neutron scattering studies, combined with bulk magnetic measurements, revealed the formation of a helical spin spiral below the critical temperature \tc~of 13 K in Cr$_{0.81}$Mn$_{0.19}$Ge, coexisting with a spin-glass state below 8 K \cite{sato1994itinerant,sato1999magnetization}. However, a more recent study by Zeng et al. challenged the spin-glass behavior in \crmnge~based on frequency-dependent AC susceptibility measurements \cite{zeng2021low}. Additionally, the authors observed a low-field topological Hall effect -- a characteristic of magnetic skyrmions—and revisited the temperature vs. magnetic field phase diagram \cite{zeng2021low}.

The helimagnetic ground state and the field-induced topological Hall effect strongly suggest the emergence of the skyrmion lattice in \crmnge. Nevertheless, the microscopic nature of the field-induced state, specifically, whether it exhibits a hexagonal SkL typical of MnSi or an HL similar to MnGe, remained unknown. Here, a detailed small-angle neutron scattering (SANS) study covering the phase diagram of \crmnge~alloy is reported. SANS data uniquely enables the assignment of magnetic phases to helical, conical, and hexagonal SkL states. Notably, \crmnge~stands out among $B20$ compounds by exhibiting a significant decrease in the helical pitch from 40\,nm to 35\,nm upon cooling from \tc~to 2 K. The low-field SkL in \crmnge~is stable at 10-30 mT and within a relatively wide temperature range of 6 K ($\sim T_\textrm{C}/2 < T < T_\textrm{C}$, $T_\textrm{C}=13$ K) compared to other bulk $B20$ compounds, where the SkL phase (or so-called $A-$phase) extends to only a few percent of \tc \cite{muhlbauer2009skyrmion,moskvin2013complex}. Furthermore, we demonstrate the alignment of a hexagonal single-domain SkL from an as-created orientationally disordered state in the polycrystal by rotating the sample in a magnetic field. Finally, we show that the facile creation of a metastable SkL achieved through a field cooling procedure that persists at the base temperature of 2\,K, even at zero applied field. This observation raises the question on how the presence of magnetic defects in the $B20$ structure promotes the stability of zero-field skyrmions.

\begin{figure*}
\includegraphics[width=1\linewidth]{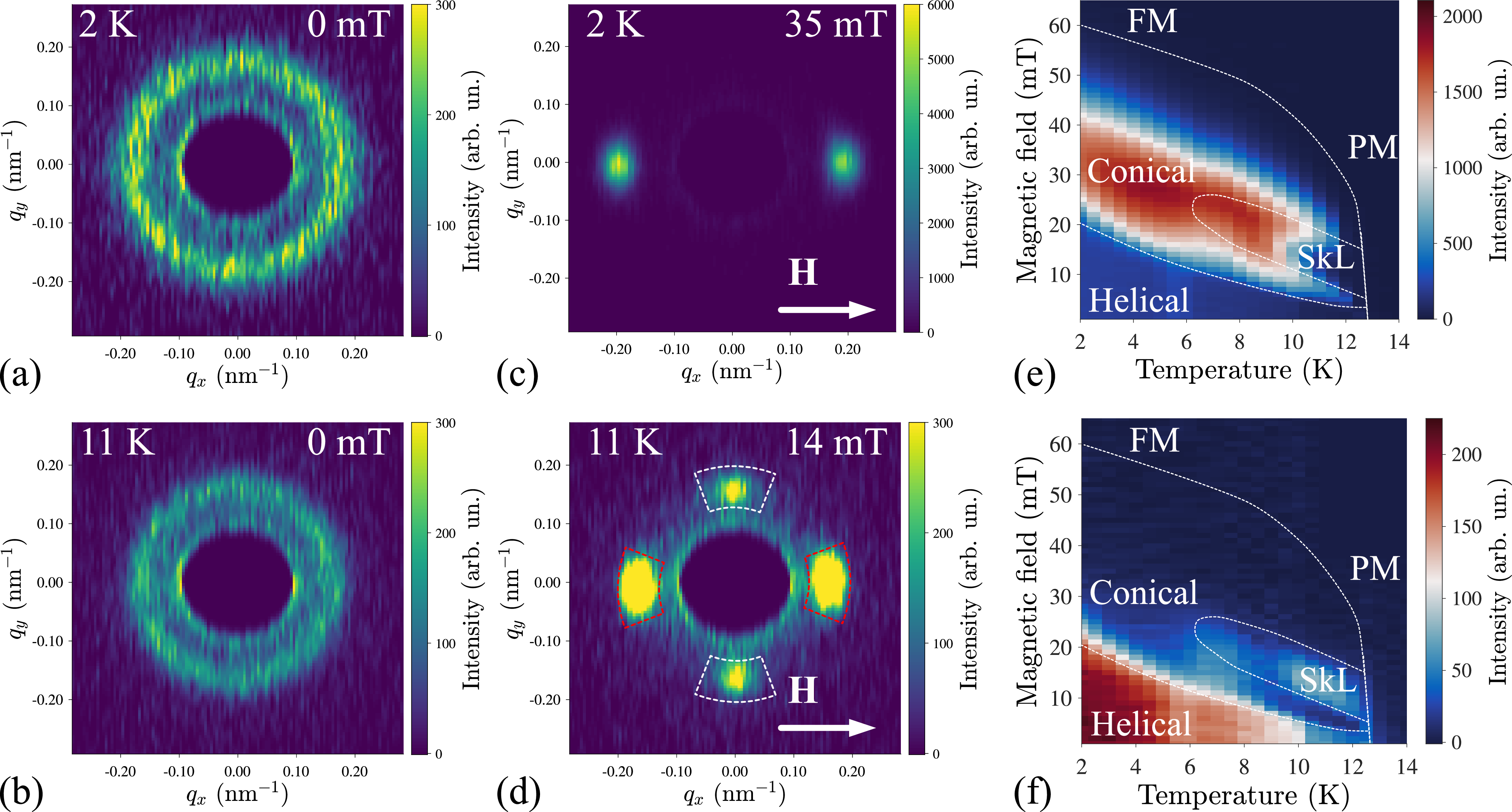}
        \caption{Ring-like zero-field small-angle neutron scattering (SANS) patterns measured at (a) 2 K and (b) 11 K in the helical state. (c) SANS in the field-induced conical state at 2 K and 35 mT. (d) SANS pattern at 11 K and 14 mT. The intensity in red and white sector boxes corresponds to the conical and skyrmion lattice (SkL) states, respectively. Magnetic phase diagrams plotted using the integrated SANS intensity in (e) red and (f) white sector boxes shown in the previous panel.}
        \label{fig1}
\end{figure*}

\section*{Results}

\textbf{Magnetic phase diagram} 

The magnetic phase diagram of our \crmnge~polycrystalline ingot was
firstly characterized by means of ac-susceptibility ($\chi'$) measurement. In agreement with prior reports, magnetic ordering initiates at \tc~of 13 K (Fig. \ref{fig0}a). Below \tc~the $\chi'(T)$ signal shows no frequency dependence (Fig. \ref{fig0}a). This result is consistent with the observations by Zeng et al. \cite{zeng2021low}, and does not support the existence of a spin glass transition in our sample. The previously reported spin-glass behavior by Sato et al. \cite{sato1999magnetic} might have originated from an amorphous impurity \cite{sato1988magnetic2}.

The SkL phase is typically suggested as a dip in the magnetic field dependence of the ac-susceptibility, embedded into conical states \cite{bauer2010quantum,bannenberg2016magnetic,karube2020metastable}. In \crmnge~the putative SkL state, or so-called $A$-phase, emerges just below \tc~in a low applied magnetic field of only 10-20\,mT (Fig. \ref{fig0}b,c). The topological Hall effect with the magnitude of $\sim 70$\,n$\Omega$cm arising from a topologically non-trivial spin texture emerges in the $A-$phase, as shown in the Supplementary Information (Figure S1).

Magnetic structures across the phase diagram have been studied in detail using SANS in a transverse magnetic field geometry. This geometry enables the unambiguous distinction between helical, conical, and SkL states. No elastic SANS signal is observed in paramagnetic (PM) and field-induced ferromagnetic (FM) phases due to the absence of incommensurate long-range magnetic order in these states. The zero-field helical state manifests as a ring pattern in SANS (Fig. \ref{fig1}a,b) owing to the orientational disorder of the crystal grains in the ingot. The application of a finite magnetic field concentrates the SANS intensity into two Bragg spots, indicative of the conical state with the wavevector oriented along the in-plane magnetic field (Fig. \ref{fig1}c). At higher temperatures, additional field-induced peaks emerge in the direction perpendicular to the conical $\mathbf{q}$-vector, representing the side-view of a SkL aligned with vortex cores antiparallel to the field (Fig. \ref{fig1}d). To determine the phase diagram, we conducted SANS measurements at selected temperatures during field ramping following a zero-field cooling procedure from a temperature well above the Curie temperature ($\sim 30$\,K). Additionally, a magnet degaussing procedure was performed at 30\,K before each measurement to eliminate possible effects of field cooling caused by residual magnetic fields trapped in the superconducting magnet’s coils. As a result, we can confidently assert that the measured magnetic phase diagram represents thermally equilibrium states. Correspondingly, the complete magnetic phase diagram of \crmnge~is constructed from the intensities of the corresponding helical, conical, and SkL Bragg peaks (Fig. \ref{fig1}e,f). The corresponding intensity vs. magnetic field plots are presented in Figure S2 of the Supplementary Material.

Remarkably, according to SANS, the SkL pocket width of 6\,K extends unusually wide compared to other cubic skyrmion hosts, with Fe$_{x}$Co$_{1-x}$Si \cite{munzer2010skyrmion} and Ag-doped Cu$_2$OSeO$_3$ \cite{neves2020effect} being a notable exceptions. Moreover, the helical-to-conical transition field in \crmnge~increases toward low temperatures, indicating increasing anisotropy in the system on cooling \cite{grigoriev2015spiral}.

\textbf{Helical wavevector} 

\begin{figure*}
\includegraphics[width=1\linewidth]{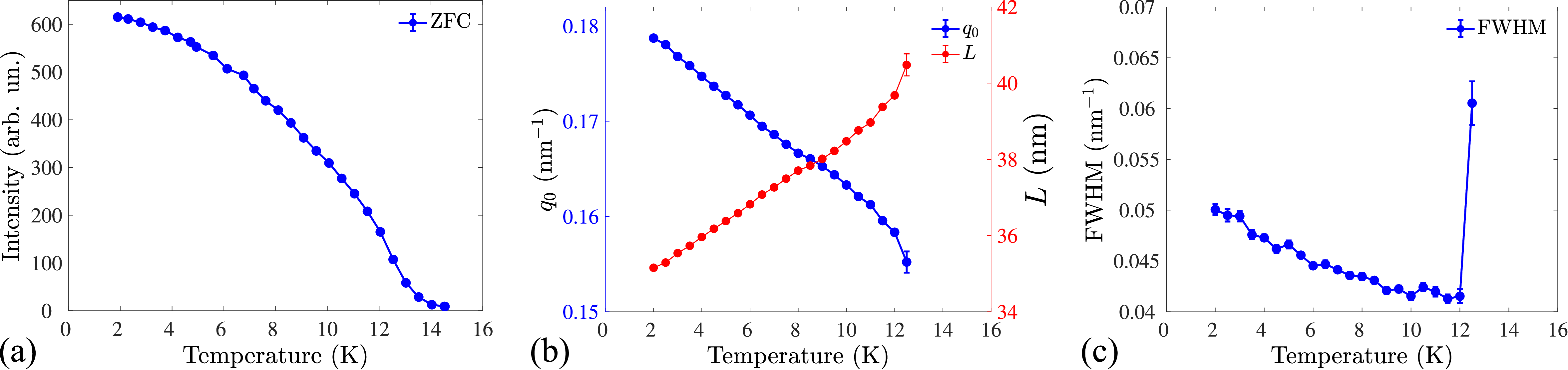}
        \caption{(a) Integrated SANS intensity on zero-field cooling. (b) Temperature dependence of the helical wavevector $q_0$ and real-space pitch $L$. (c) Full-width at half maximum (FWHM) of the helical peak as a function of temperature.}
        \label{fig2}
\end{figure*}

Upon the development of the helimagnetic order from the paramagnetic state (Fig. \ref{fig2}a) an intriguing feature is observed in \crmnge~in the temperature dependence of the spiral wavevector $q_0$ and its periodicity $L$, respectively (Fig. \ref{fig2}b). In cubic chiral magnets, $L$ is determined by the ratio between exchange interaction and DMI constants $J$ and $D$: $L \sim J/D$ according to the Bak-Jensen model \cite{bak1980theory}. 

Unlike the typical decrease observed in single crystals of MnSi \cite{grigoriev2006magnetic} or Cu$_2$OSeO$_3$ \cite{chacon2018observation} or a slight increase observed in FeGe \cite{lebech1989magnetic} and Mn$_{1-x}$Co$_x$Ge \cite{altynbaev2018magnetic} samples, \crmnge~exhibits a substantial increase in $q_0$ of 15\% upon cooling, which is similar to Fe$_{1-x}$Co$_x$Si \cite{grigoriev2007magnetic,bannenberg2016extended}. Such a change can be attributed to the contribution of anisotropic exchange interaction (AEI) \cite{baral2023direct}. Moreover, the increase in helical-to-conical transition field and the almost linear change in $q_0 (T)$ suggest a gradual enhancement of the AEI in \crmnge~at low temperatures. Simultaneously, with the decrease in the helical pitch, its correlation length decreases with temperature, evident from the increase in the full-width at half maximum (FWHM) of the peak (Fig. \ref{fig2}c). Future high-resolution resonant small-angle x-ray scattering studies can directly measure the temperature dependence of the AEI constant \cite{baral2023direct}.

Another family of chiral magnetic materials that also exhibits an increase in helical $q_0 (T)$ and FWHM at low temperatures is $\beta$-Mn-type alloys Co-Zn-Mn \cite{karube2020metastable} and molybdenum nitride FePtMo$_3$N \cite{sukhanov2020robust}, with the exact mechanism responsible for this change remaining unknown, but connected to the magnetic frustration effect in the $\beta$-Mn lattice  \cite{ukleev2019element,ukleev2021frustration,ukleev2022spin}. Further studies aiming at determining the microscopic nature of the interactions are needed to determine the importance of any magnetic frustration in \crmnge. 

\textbf{Skyrmion lattice ordering} 

While the transverse-field SANS geometry is effective for mapping magnetic phase diagrams, it does not distinguish between hexagonal SkL and three-dimensional HL. To address this, we employed an experimental algorithm proposed by Gilbert et al. \cite{gilbert2019precipitating} to anneal two-dimensional SkL structures with improved orientational order from the as-created skyrmion lattices with full orientational-disorder expected in polycrystal samples.

\begin{figure*}
\includegraphics[width=1\linewidth]{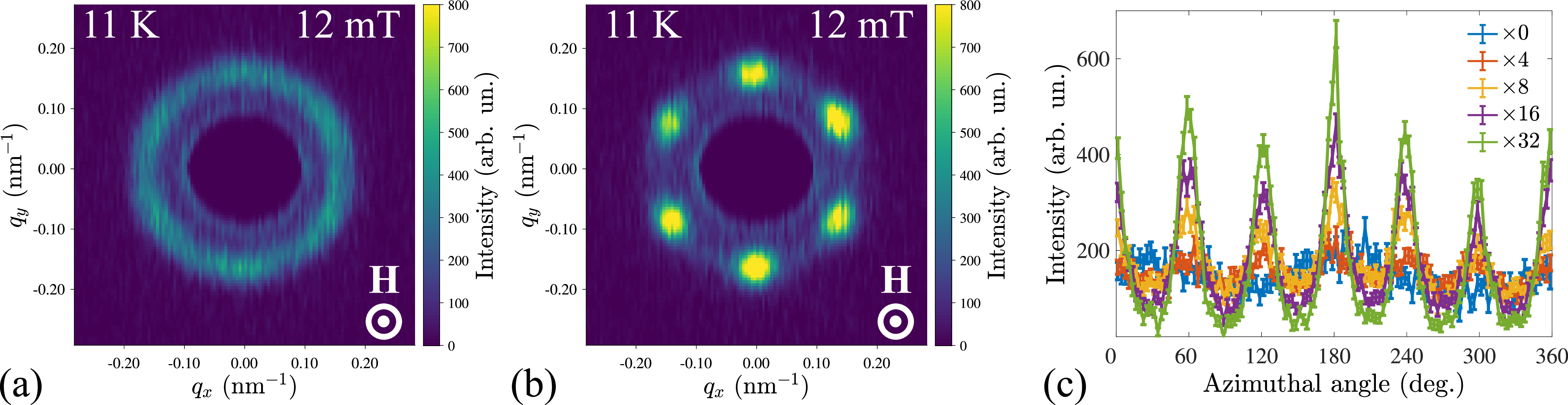}
        \caption{(a) Ring-like SANS pattern measured at 11 K and 12 mT (SkL phase) in the longitudinal-field geometry. (b) Six Bragg peaks from the ordered hexagonal SkL emerging in SANS after rotating the sample in field. (c) Azimuthal distribution of the radially integrated SANS intensity after different rotation cycles.}
        \label{fig3}
\end{figure*}

The SANS pattern from the zero-field cooled untrained \crmnge~sample in the longitudinal geometry at 11\,K and 12\,mT initially exhibits a ring-like pattern (Fig. \ref{fig3}a). After a series of in-field sample rotations by $\pm20^\circ$, the SANS intensity has concentrated into six spots representing the hexagonal SkL with well-defined orientational order (Fig. \ref{fig3}b). The possible effect of crystal grains reorientation in excluded in the present case owing the solid nature of the polycrystalline pellet. Therefore, the single-domain SkL order emerges from interactions between individual grains \cite{gilbert2019precipitating}. The training effect reaches its saturation after ca. 28 rotations as suggested from the azimuthal profiles of the radially integrated SANS patterns (Fig. \ref{fig3}c). Additionally, a rocking SANS scan was carried out on the field-training-ordered SkL state. The FWHM of the SkL SANS peak along the neutrons path is of 7.5$\circ$ which is comparable to $B20$ single crystals with large mosaicity \cite{gilbert2019precipitating}.

The result of the training distinctly demonstrates that, in contrast to MnGe, which hosts a cubic HL, \crmnge~hosts a conventional quasi-2D hexagonal SkL similar to prototype $B20$ compounds like MnSi and FeGe.

\textbf{Metastable skyrmion lattice} 

Finally, we demonstrate the metastability of the SkL at low temperatures. Conventional field cooling through the $A$-phase often results in a metastable (supercooled) SkL state in chiral magnets with chemical disorder \cite{milde2013unwinding,karube2020metastable,wilson2019measuring,ukleev2022topological}. Given the significant doping by Mn, a metastable SkL in \crmnge~could be anticipated. Indeed, a field cooling at 12\,mT from the ordered SkL state with a cooling rate of 3\,K/min results in the well-preserved SkL state at 2\,K, i.e., far outside of thermally equilibrium SkL phase (Fig. \ref{fig4}a). The crystalline quality of the field-ordered SkL is similar to the one of the helical state at both 11\,K and 2\,K, and is likely limited by the presence of the magnetic defects. Remarkably, the metastable phase survives even at zero field (Fig. \ref{fig4}b) and is robust against magnetic field perturbations from both positive and negative magnetic fields (Fig. \ref{fig4}c). Unlike metastable skyrmions in MnSi \cite{nakajima2017skyrmion}, Cu$_2$OSeO$_3$ thin plate \cite{takagi2020particle}, or Co$_9$Zn$_9$Mn$_2$ \cite{karube2017skyrmion} the hexagonal SkL in \crmnge~does not transform into a square one on a field decrease.

\section*{Discussion}

Our comprehensive study of the $B20$-type chiral magnet \crmnge~has revealed intriguing magnetic behaviors distinct from the parent compounds, CrGe and MnGe.

The constructed magnetic phase diagram, utilizing transverse-field SANS geometry, allowed us to differentiate between helical, conical, and SkL states. Notably, \crmnge~exhibits an unusually wide temperature pocket of 6 K ($\sim T_\textrm{C}/2$) for the low-field induced SkL, extending from just below \tc~in a magnetic field as low as 10-20 mT. The helical-to-conical transition field increases towards lower temperatures, indicating the magnetic anisotropy enhances in magnitude on cooling in the system. The temperature dependence of the spiral wavevector suggests a significant increase in exchange anisotropy at low temperatures. Moreover, the simultaneous decrease of the spiral coherence length (growth of FWHM in Fig. \ref{fig2}b) and increase of AEI, suggests that \crmnge~can be viewed as a model system of disordered helimagnet~\cite{utesov2015,utesov2019} with random anisotropy field due to local spatial inhomogeneities of Mn and Cr concentrations. It provides a natural description of the elastic Bragg peaks broadening in the mixed $B20$ helimagnets and, in the case of a strong disorder, of the transition to the phase with a short-range chiral order previously discussed for Mn$_{1-x}$Fe$_x$Si in Ref.~\onlinecite{grigoriev2021critical}. A detailed study along these lines is out of the scope of the present paper and will be presented elsewhere.

Additionally, by utilizing a sample rotation algorithm, we confirmed the presence of a hexagonal SkL in \crmnge, distinguishing it from MnGe-based $B20$-type compounds that host three-dimensional HLs. This result underscores the versatility of the manganese-germanide derivative compounds in hosting diverse topological spin textures. Furthermore, our investigation into the metastability of the SkL at low temperatures revealed its persistence even at zero applied field. The periodicity of the SkL decreases after the field cooling, following the same tendency at the helical state. This robust metastable state with flexible size of the skyrmion adds another layer to the intriguing magnetic behavior of \crmnge. 

In summary, our study not only enhances our understanding of the magnetic phases in \crmnge~and proves microscopically that the origin of the topological Hall effect is associated with skyrmion lattice order, it also contributes valuable insights into the rich magnetic phase landscape of $B20$-type chiral magnets. The observed features, such as the wide temperature range for the low-field induced SkL and the enhanced exchange anisotropy, make \crmnge~a compelling system for further exploration. Furthermore, theory proposes CrGe as a Weyl semimetal \cite{klotz2019electronic}, which provides tantalising suggestions that the Mn-doped compounds are a fruitful platform for exploring the interplay between electronic and magnetic topologies.

\section*{Methods}
\textbf{Sample synthesis.} Polycrystalline ingots of \crmnge~were synthesized using the arc-melting technique, followed by annealing according to the procedure outlined in Ref. \onlinecite{zeng2021low}. The constituent elements (Mn, Cr and Ge) of purity better than 99.9\%
were used to prepare the sample. The sample was melted five times by turning upside down. 3wt\% extra Mn was added during melting to adjust the weight loss of Mn and maintain the stoichiometry of the sample. The as-cast sample was sealed inside evacuated quartz tubes under partially
filled Ar atmosphere ($P\sim200$\,mbar) and annealed at 900$^\circ$C for 7 days followed by furnace cooling. The annealed sample was crushed into powder and then compacted in
cylindrical shape using a hydraulic press. The compacted sample annealed again for additional 7 days at 900$^\circ$C followed by quenching in cold water.

\textbf{Magnetization and magnetotransport measurements.} The ac-susceptibility and dc-magnetization measurements were carried out using a Quantum Design MPMS 3 magnetometer. The ac-susceptibility measurements were performed for the temperature interval of 2--50 K by varying the frequency from 1 to 750 Hz with a constant excitation field of 0.3\,mT. Field dependence of the ac-susceptibility data were collected for an excitation field of 0.3 mT at a frequency of 200 Hz with applied magnetic fields up to 600 mT. The isothermal dc-magnetization measurements were performed within the temperature interval of 2--15 K and in applied magnetic fields up to 1 T. The magnetotransport measurements were executed using the Electrical Transport Option (ETO) option of the MPMS 3 magnetometer. A bar-shaped sample of approximate dimensions $5.7\times2.3\times0.5$ mm$^3$ was used for magnetoresistance and Hall resistance measurements for the temperature range of 10--14 K with applied magnetic fields up to 1 T.

\begin{figure*}
\includegraphics[width=1\linewidth]{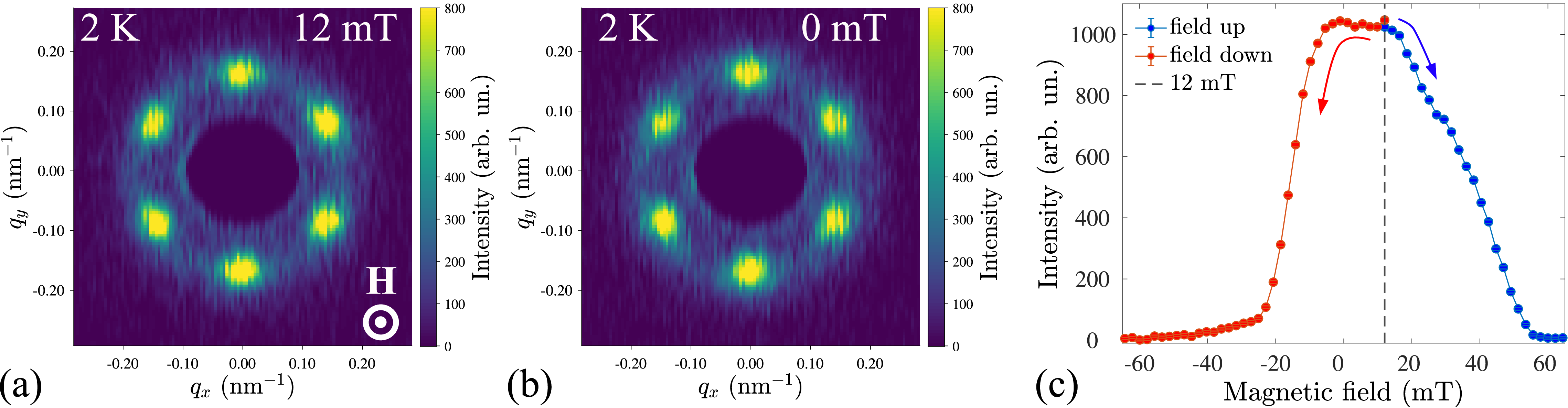}
        \caption{SANS patterns of the metastable SkL at 2 K and (a) 12 mT and (b) 0 mT after field cooling. (c) Magnetic field dependence of the integrated SANS intensity from the metastable SkL.}
        \label{fig4}
\end{figure*}

\textbf{Neutron scattering.} Small-angle neutron scattering (SANS) measurements were conducted at the Paul Scherrer Institut (PSI) using the SANS-I instrument. Neutrons with a wavelength of 10\,\AA, were employed, and the experimental setup included a sample-detector distance of 10 m and collimation of 8 m. Magnetic field control was achieved using a 6.8 T horizontal-field cryomagnet (MA7) with $\pm22.5^\circ$ windows, enabling SANS experiments in both longitudinal ($\mu_0 H$ parallel to the incident neutron wavevector $\mathbf{k}_i$) and transverse magnetic field $\mu_0 H$ perpendicular to $\mathbf{k}_i$ geometries. Additionally, the sample could be rotated around the vertical axis inside the magnet. Due to the disc-like shape of the sample the demagnetization factors are different between the transverse and longitudinal field SANS geometries. Therefore, to correct for this difference, the magnetic field values for the longitudinal field SANS measurements (Figs. \ref{fig3},\ref{fig4}) are divided by 1.83. Background SANS signals were measured in the paramagnetic state of the sample at 20 K and subsequently subtracted. The analysis of all SANS data was performed using the GRASP software \cite{dewhurst2023graphical}. 

\section*{Data availability}

The data is available from the corresponding author upon request.

\section*{Acknowledgements}

This work is based partly on experiments performed at the Swiss spallation neutron source SINQ, Paul Scherrer Institute, Villigen, Switzerland. The SANS experiments were carried out using SANS-I instrument a as a part of the proposal 20230352. We kindly acknowledge x-ray and correlative microscopy and Spectroscopy CoreLabs of HZB for provided instrumentation and R. Gunder anf F. Ruske for their assistance with the sample characterization. J.S.W. acknowledges funding from the SNSF Project 200021\_188707. O. I. U. acknowledges financial support from the Institute for Basic Science (IBS) in the Republic of Korea through Project No. IBS-R024-D1.

\section*{Author contributions}
T.S. and L.C. synthesized and characterized the samples; V.U. and J.S.W. performed the neutron measurements; V.U., O.I.U. and J.S.W. analyzed the data and wrote the manuscript. V.U. and L.C. jointly conceived the project.

\section*{Competing interest}
The authors declare no competing interests.


\bibliography{biblio}

\end{document}